\documentclass{svjour3}                     % onecolumn (standard format)
\smartqed  % flush right qed marks, e.g. at end of proof
\usepackage{graphicx,amsmath,amssymb,amsfonts}
\usepackage{epstopdf}

\newcommand{\ket}[1]{\left|#1\right\rangle}

\begin{document}

\title{Interband transitions and interference effects in superconducting qubits
\thanks{This work was financially supported by the Academy of Finland, the Finnish Cultural Foundation, the Magnus Ehrnrooth Foundation, the Vaisala Foundation of the Finnish Academy of Science and Letters, EU-INTAS 05-10000008-7923, the Dynasty foundation, the presidential grant MD-4092.2007.2, and the EC-funded ULTI Project (Contract RITA-CT-2003-505313)}
%Grants or other notes
}


\author{Antti Paila\and
        Jani Tuorila \and
        Mika Sillanp\"a\"a \and
         David Gunnarsson \and
         Jayanta Sarkar \and
         Yuriy Makhlin \and
         Erkki Thuneberg \and
         Pertti Hakonen
         %etc.
}

\authorrunning{A. Paila et al.} % if too long for running head

\institute{Antti Paila, Mika Sillanp\"a\"a, David Gunnarsson,
Jayanta Sarkar, and Pertti Hakonen \at
              Low Temperature Laboratory, Helsinki University of Technology, FI-02015 TKK, Finland \\
              Tel.: +358-9-4512964\\
              Fax:  +358-9-4512969\\
              \email{pjh@boojum.hut.fi}           %  \\
           \and
            Jani Tuorila and Erkki Thuneberg
            \at Department of Physical Sciences, University of Oulu,  FI-90014, Finland
            \and
           Yuriy Makhlin \at
              Landau Institute for Theoretical Physics, 119334 Moscow, Russia
}

\date{Received: date / Accepted: date}

\maketitle

\begin{abstract}
We investigate phase-sensitive interference effects in a
periodically  $\sin(2\pi f_{\rm rf} t)$-driven, artificial two-state
system connected to a microwave resonator at $f_{LC} \simeq 800$ MHz.
We observe two kinds of multiphoton transitions in the two-state system, accompanied by: 1) Several quanta
from the drive at $f_{\rm rf}$ and 2) one quantum at $f_{\rm rf}$ and
several at $f_{LC}$. The former are described using phase-sensitive
Landau-Zener transitions, while the latter are discussed in terms of
vibronic transitions in diatomic molecules. Interference effects in
the vibronic transitions governed by Franck-Condon coefficients are
also considered.

\end{abstract}

\section{Introduction}

Interference effects have been found to play an important role in
the dynamics of qubits
\cite{shytovEPJB03,Shevchenko05,MIT,LZprl,Berns2006,WilsonPRL07}. For their
understanding, the interference phenomena can be considered from
different points of view: as interference between successive
Landau-Zener (LZ) tunneling events
\cite{Landau32,Zener32,Stueckelberg32}, analogous to % optical
Mach-Zehnder interference \cite{Sprinzak}, or in terms of spin
dynamics \cite{Majorana32}.
% using Bloch equations \cite{Bloch46,Abragam}.
In charge-phase qubits, strongly coupled to
a microwave resonator, analogy with vibrational transitions in
diatomic molecules has been pointed out \cite{Gunnarsson}. The fact that the latter
phenomena are related to interference effects has its origin in the phase-space dynamics \cite{Wheeler}.

The state of a superconducting charge-phase qubit or a Cooper-pair
box (CPB) can be monitored continuously by measuring its reactive
response, either as an effective capacitance or inductance, which both
are due to the curvature of the qubit's energy bands with respect to
charge or phase degrees of freedom. When such a system is made as a
part of an electric $LC$-oscillator circuit, any change of reactance,
caused by the evolution of occupancies of the qubit's energy levels,
will shift the resonance frequency of the resonator. This scheme has
been used in several recent experiments, typically having the qubit,
with splitting $\Delta E$, connected to a coplanar stripline
cavity near resonance \cite{CQED}. We have, however, worked in
the fully detuned limit where $f_{LC}\ll \Delta E/h$. At weak drive,
such a read-out scheme perturbs the investigated system only
weakly, allowing studies of inherent interference phenomena of the
periodically driven two-level system.

\begin{figure}
\begin{center}

\includegraphics[width=10.0cm]{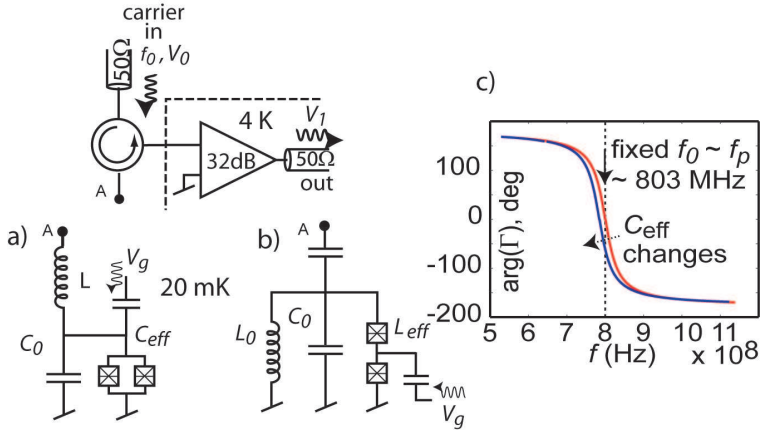} \caption{Schematics of our experimental
configurations: a) Cooper-pair box and b) the inductively read
charge-phase qubit (LSET). In both configurations the qubit-induced
change in the "quantum" reactance alters the resonant frequency
$\omega_{LC}/2\pi$ of the tank circuit. This change causes a shift
in phase of the reflected, constant-frequency microwave signal as
illustrated in c). In the illustration, $C_{\rm eff}$ grows which leads
to a reduction of $\omega_{LC}$ and, consequently, to a  decrease of
the reflection phase arg($\Gamma$). \label{fig:Setup}}
\end{center}
\end{figure}

We have investigated interference effects in the Cooper-pair box and
in a charge-phase qubit circuit, configured as dual to the CPB (i.e., connected to the resonator via the phase, rather than the charge port), see Fig.\  \ref{fig:Setup}.
%By dual behavior we mean here that in these circuits the behavior with
%respect to charge and phase variable are of equal character.
Both circuits provide good model systems of interference effects in
periodically driven qubits. In the latter case, a clearly
stronger coupling between the qubit and the resonator could be
achieved, which made a difference in the observed effects.

The interference effects in these two circuits are
diverse mostly due to the difference in the coupling between the
resonator and the qubit. On the whole, interference phenomena can be
understood in terms of multiphoton transitions. In strongly driven qubits,
clear multiphoton transitions have been observed
when the energy quantum of the rf-drive, multiplied by a small integer,
matches the qubit level splitting \cite{NakamuraMultiphoton,Saito04}.
In fact,
multiphoton transitions allow a clear-cut distinction between
our main findings. We may classify our observations as transitions where: 1) Several quanta
from the drive at $f_{\rm rf}$ and 2) one quantum at $f_{\rm rf}$ and
several at $f_{LC}$ are exchanged between the qubit and its surrounding entities,
i.e. the classical rf-drive port and the $LC$ resonator.
The first class can be treated as a sequence of phase-sensitive
Landau-Zener transitions which produce a characteristic
interference pattern, dependent on the "finesse" of the interferometer,
namely the decoherence of the system. The latter
class of transitions can be described in terms of vibronic
transitions in diatomic molecules where potential changes are fast
compared to the vibrational frequencies and non-adiabatic coupling
between levels differing by a large number of quanta becomes
possible. These phenomena include creation of several
resonator quanta from the ground state, which is the non-adiabatic
behavior referred to as the dynamical Casimir effect
\cite{Casimir}.

In this paper, we summarize some of our recent results \cite{LZprl,Gunnarsson} on interference effects in a driven superconducting charge qubit coupled to a resonator. The original references, in particular, used theoretical considerations and numerical simulations of the Bloch equations and the linear-response approximation for a detailed analysis of the shape of interference fringes when
dissipation is important. Instead, here we try to concentrate on
robust features which are amenable to general conclusions. For
example, part of our analysis will be dealing with the phenomenon of destruction of tunneling.

The paper is organized as follows: First, we start with basics of the
Cooper-pair box  in Sect. 2 and describe the dynamics of a two-level
system in Sect. \ref{s.l}, paying special attention to the destruction of
tunneling from the ground state to the excited state. Sect.~\ref{s.m} covers our experimental work and results on CPBs, and it includes some considerations aiming towards understanding of the Stokes phase and its relation to geometric phases
in a spin-1/2 system. We present data
at phase bias $\pi$ (minimum band gap), and we
analyze, especially, data in the regime of destruction of tunneling at
the charge degeneracy. Results at zero phase bias can be found in Ref. \cite{LZprl}.
Section 5 describes the analogy of spin flips
in strongly coupled qubits with vibronic transitions in diatomic
molecules. We do not describe much our experimental
techniques, but to make the paper self-contained, we have explained
some of the main technical points when presenting the experimental
results.
% in an Appendix.

\section{Cooper-pair box}

A split Cooper-pair box (CPB) is formed by a single-Cooper-pair
transistor (SCPT) embedded into a small superconducting loop
\cite{Bouchiat98,Nakamura99,MSS}, see Fig.\ \ref{fig:Setup}. The charging energy of the
CPB, $E_{\rm c} = e^2/2C_\Sigma \sim 1$ Kelvin, is  given by the total
capacitance $C_\Sigma$ which includes the junction capacitances, the gate
capacitance $C_{\rm g}$, and the self-capacitance of the island. The
effective Josephson energy is given by the sum of the energies of
the individual junctions  $(E_{{\rm J}1}+E_{{\rm J}2}) \cos(\phi/2)=E_{\rm J}
\cos(\phi/2)$, which is tunable by magnetic flux $\Phi $, \textit{
i.e.} by the superconducting phase across the two junctions, $\phi =
2 \pi \Phi / \Phi_0$. Here, $\Phi_0 = h/2e$ is the superconducting magnetic flux
quantum.

We may write the Hamiltonian in the charge basis as
\begin{eqnarray}
\hat{H}&=& \sum_n \bigg[E_{\rm c}\left(\hat{n}-n_{\rm g}\right)^2 |n\rangle\langle n|
-\frac{E_{\rm J}}2\cos{{\frac\phi2}} \left(|n-2\rangle\langle n|+|n+2\rangle\langle n|
\right) \nonumber \\
&&+i\frac{E_{\rm J}d}2\sin{\frac\phi2}\left(|n-2\rangle\langle n|-|n+2\rangle\langle
n| \right)\bigg].
\end{eqnarray}
%DEFINITIONS
Here $\hat n$ denotes the number of extra electron charges on the
island, and $n_{\rm g} = C_{\rm g} V_{\rm g} / e$ is the charge in electron units induced by the gate voltage $V_{\rm g}$ on the gate capacitor with capacitance $C_{\rm g}$.
 $\hat n$ is conjugate to $\hat\theta/2$, where $\hat\theta$ is
the superconducting phase on the island. The asymmetry of the two Josephson
 junctions of the CPB is described by
 $d=(E_{{\rm J}1}-E_{{\rm J}2})/E_{\rm J}$.

\begin{figure}
\begin{center}
\includegraphics[width=5.0cm]{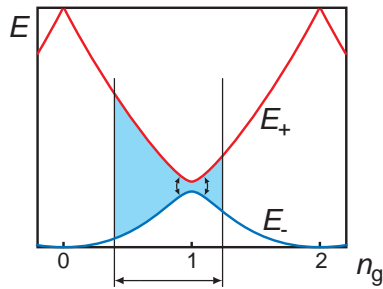} \caption{The two lowest energy bands of Cooper-pair box as a function of the gate charge $n_{\rm g}$ (in units of $e$). In sweeping $n_{\rm g}$ (horizontal arrows), Landau-Zener tunneling between the bands takes place close to the point where the energy difference is at minimum (curved arrows). Successive sweeps can lead to interference of the tunneling events. The interference depends on the phase (\ref{dynamic}) accumulated in the shaded area.
\label{f.LZfig}}
\end{center}
\end{figure}

By assuming that $E_{\rm c} \gg E_{\rm J}$ and that $n_{\rm g}$ is close to 1, one can reduce the circuit to a two-state system, with the Hamiltonian
\begin{equation}
\hat{H}=\frac12\left(\begin{array}{cc}
-4E_{\rm c}(1-n_{\rm g}) &-E_{\rm J}[\cos{(\phi/2)}+id\sin{(\phi/2)}]  \\
-E_{\rm J}[\cos{(\phi/2)}-id\sin{(\phi/2)}] &4E_{\rm c}(1-n_{\rm g})
\end{array}\right)
\end{equation}
in the basis of the relevant charge states $|0\rangle$ and $|2\rangle$.
When this energy operator is expressed as a linear combination of
the Pauli matrices $\hat{H}=- \frac{1}{2} ( B_z \hat{\sigma}_z$ +
$B_x \hat{\sigma}_x+B_y \hat{\sigma}_y ),$ we get the magnetic field
components as $B_z=4E_{\rm c}(1-n_{\rm g})$, $B_x=E_{\rm J}\cos{(\phi/2)}$ and
$B_y=dE_{\rm J}\sin{(\phi/2)}$.
The energies of the two states as a function of $n_{\rm g}$ are illustrated in Fig. \ref{f.LZfig}.
%%%%%%%%%%%%%%%%%%%%%%%%%%%%%%%%%%
%%%%%%%%%%%%%%%%%%%%%%%%%%%%%%%%%%

In the experiments $E_{\rm J}=0.6$ K and $d=0.22$, which means that the magnitude for the
off-diagonal components $|\Delta|$ ranged over 1.4--6.2~GHz~\cite{LZprl}.
The non-linear parametric capacitance,
which has been the cornerstone of our interference studies, has been
experimentally studied in Refs. \cite{MikaQcap,Duty05}.

\section{Landau-Zener interference}\label{s.l}

We study the effect of sweeping the gate charge $n_{\rm g}$. We denote the energy eigenstates by $\ket{-}$ and $\ket{+}$, and suppose the system is initially in the lower energy state $\ket{-}$, see Fig. \ref{f.LZfig}. Two cases can be distinguished. When the sweep rate is small compared to the energy level difference, the system stays in the lower state. The alternative case is that the sweep rate is comparable to the level spacing. This may take place close to the degeneracy points, where the energy bands would cross in the absence of the Josephson coupling.  There the system can tunnel from the lower state to the upper in a process known as  Landau-Zener
tunneling. The tunneling probability  in a single sweep is given by~\cite{Landau32,Zener32,Stueckelberg32,Majorana32}
\begin{equation}\label{eq:Plz}
P_{LZ}=e^{-2\pi\gamma}, \qquad
\gamma=\frac{2\pi}{h}\frac{\Delta^2}{v}.
\end{equation}
Here $v$ is the speed at which the sweep passes the crossing point
$v=|\mathrm{d}(\epsilon_0-\epsilon_2)\large{/}\mathrm{d}t|$ and
$\epsilon_0-\epsilon_2=4E_c(1-n_g)$ is the energy difference in the absence of the Josephson coupling. In
Eq. (\ref{eq:Plz}) the adiabaticity parameter $\gamma$ defines
whether the development is adiabatic ($\gamma \gg 1 $) or
sudden ($\gamma \ll 1$).

When the system is coherent, and the degeneracy point is crossed
several times, the transition amplitudes for each subsequent pass
have to be added for evaluating the transition probability. These
amplitudes may be tracked
by the `scattering' matrix $S$, defined by
(cf.~Refs.~\cite{Zener32,GefenRev,Kayanuma97,wubs05}):
\begin{equation}\label{amplitudes}
\left(
\begin{array}{c}\ket{-}\\\ket{+}\\\end{array}
\right) \Rightarrow
%-i
\left(
\begin{array}{cc}
  \sqrt{1-P_{\mathrm{LZ}}} \exp (i \tilde\phi_S) & i\sqrt{P_{\mathrm{LZ}}} \\
  i\sqrt{P_{\mathrm{LZ}}} & \sqrt{1-P_{\mathrm{LZ}}} \exp (-i \tilde\phi_S ) \\
\end{array}%
\right) \left(
\begin{array}{c}\ket{-}\\\ket{+}\\\end{array}
\right),
\end{equation}
where phase factors are chosen to simplify the matrix.
Here the scattering phase is $\tilde\phi_S=\phi_S-\pi/2$, where the
Stokes phase $\phi_S$ depends on the adiabaticity parameter
$\gamma$:
\begin{equation}\label{gamma}
\phi_S=\pi/4+\arg(\Gamma(1-i\gamma))+\gamma(\ln \gamma -1) .
\end{equation}
In the adiabatic limit, $\tilde\phi_S \rightarrow -\pi/2$, while in
the sudden limit, $\tilde\phi_S = -\pi/4$.

Away from the degeneracy point, the eigenstates $\ket{-}$ and
$\ket{+}$ accumulate the relative dynamical phase
\begin{equation}\label{dynamic}
    \varphi_d = \varphi^{(+)} - \varphi^{(-)} = -\frac{1}{\hbar} \int \left[ E_+(n_{\rm g}(t)) -
    E_-(n_{\rm g}(t)) \right]
    dt\, .
\end{equation}
Thus, the condition for constructive interference is that $\varphi_\mathrm{L} - 2 \tilde\phi_S$ and $\varphi_\mathrm{R} - 2 \tilde\phi_S$ are multiples of $2\pi$,
where $\varphi_\mathrm{L}$ and $\varphi_\mathrm{R}$ refer to the
dynamical phases accumulated on the left and right sides of the degeneracy
point, respectively. For example, in the adiabatic limit,
$\varphi_\mathrm{L,R}$ have to be odd multiples of $\pi$.

The LZ interference can also be employed to suppress the tunneling
to the upper level \cite{Destruction}. In fact, the contrast for
destructive interference in the experimental data looks often better
than for constructive interference, and these conditions can be
employed more simply to determine the behavior of the system as the
ground-state response in the measurement is well known. In this case
the interference conditions are
\begin{equation}\label{phasecondition}
\varphi_\mathrm{L} + 2\pi \ell = 2 \phi_S\ {\rm and}\ \varphi_\mathrm{R} + 2\pi \ell = 2 \phi_S
\end{equation}
with an integer $\ell$.

In the sudden limit, $\gamma \ll 1$, the
periodically driven two-level system can be solved in a rather
straightforward manner \cite{Kayanuma94}. At degeneracy, the probability of being
in the upper state varies periodically with time according to
\begin{equation}\label{bessel}
P(t)=\sin^2\left[B_x J_0(A/\hbar\omega_{\rm rf})\Delta t /\hbar\right]\,,
\end{equation}
where $J_0$ denotes the zeroth Bessel function, and $A$ ($=4E_{\rm c}
n_{\rm g}^{\rm rf}$ below) is the drive amplitude at frequency
$f_{\rm rf}=\omega_{\rm rf}/2\pi$. One notices that the destruction of tunneling
takes place, when the ratio $A/\hbar\omega_{\rm rf}$ coincides with a
zero of $J_0$.

The description of interference patterns in a Cooper-pair box using the
scattering matrix formalism has been discussed in Refs.~\cite{YuriyLammi,Vietnam}. In general, relaxation phenomena should
be included as they have strong influence on the sharpness of the
interference fringes. For this reason, our main analysis method has
been based on spin-1/2 NMR simulations \cite{LZprl,Vietnam}, which
also includes linear response calculations in order to obtain the
measured, effective capacitance. Alternatively, the dressed-state
approach may be employed as has been done by C. Wilson \emph{et al.}
\cite{WilsonPRL07}.

\section{Measurement results on CPB}\label{s.m}

We have performed low-dissipation microwave reflection
measurements \cite{LSET,LeifPRB05,MikaThesis} on a series $LC$
resonator in which the box effective capacitance,  $$C_{\mathrm{eff}}^{\pm} = -
\frac{\partial^2 E_{\pm}(\phi, n_{\rm g})}{\partial V_{\rm g}^2} = -
\frac{C_{\rm g}^2}{e^2} \frac{\partial^2 E_{\pm} (\phi, n_{\rm g})}{\partial
n_{\rm g}^2},$$ is a part of the total capacitance
$C_S+C_{\mathrm{eff}}^{\pm}$, where the superscript $\pm$ refers to ground and
exited states of the qubit.\footnote{For a more detailed description of $C_{\mathrm{eff}}$, see Refs. \cite{MikaQcap,LZprl}}
The resonator is formed by a surface mount inductor of
$L = 160$ nH. With a stray capacitance of $C_S = 250$ fF due to the
fairly big lumped resonator, the resonant frequency is $f_0 = 800$
MHz, and the quality factor $Q \simeq 16$ is limited by the external
$Z_0 = 50 \ \mathrm{\Omega}$.  When $C_{\mathrm{eff}}^{\pm}$ varies, the
phase and amplitude of the reflected signal
$V_{\mathrm{out}}=\Gamma V_{\mathrm{in}}$ change, which is measured
by the reflection coefficient $\Gamma = (Z-Z_0)/(Z+Z_0)=\Gamma_0
e^{i \arg (\Gamma)}$. Here, $Z$ is the resonator impedance as seen
from the end of the  $ 50\ \mathrm{\Omega}$ coaxial cable used for the
reflection measurement. The variation in $\arg (\Gamma)$ due to
modulation in $C_{\mathrm{eff}}^{\pm}$ is up to 40$^{\circ}$ in our
measurements, corresponding to a shift of resonance frequency
$\Delta f_p \simeq 6$~MHz.
%In addition to band pass filtering, we
%used two circulators at 20~mK to prevent the back-action noise of
%our cryogenic low-noise amplifier from reaching the qubit.
In all the measurements, the weak probing signal
$V_{\mathrm{in}}$ at frequency $f_{\rm m}$ was continuously
applied, in addition to the DC-bias and the rf-drive. Thus, the total gate charge variation, in units of $e$, can be written as
$n_{\rm g}(t)= n_{{\rm g}0} + n_{\rm g}^{\rm rf} \sin(2\pi f_{\rm rf} t) + n_{\rm g}^{m} \sin(2\pi f_{\rm m} t)$, where the $1^{st}$, $2^{nd}$, and $3^{rd}$ terms correspond to the DC, rf, and measurement drives, respectively.

\begin{figure}
\begin{center}
\includegraphics[width=12.0cm]{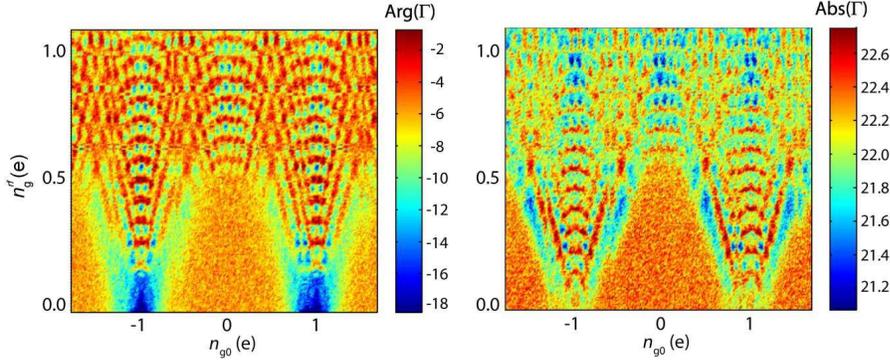} \caption{Reflection
phase (left-a) and magnitude (right-b) measured in the configuration of
Fig. 1a at rf-drive frequency of $f_{\rm rf} = 4$ GHz using phase bias
$\phi = \pi$, corresponding to the level repulsion of $2 \Delta =
E_{\rm J}d= 2.7$ GHz. $n_{\rm g}^{\rm rf}$ specifies the amplitude of the rf drive
in electrons and $n_{{\rm g}0}$ is the DC charge bias. The vertical bars
give the color scales in degrees and dB for the left and right
frames, respectively.\label{fig:PiFlux}}
\end{center}
\end{figure}

We have made extensive scans of the CPB reflection by varying the LZ
drive frequency $f_{\rm rf}=0.1$--20~GHz and its amplitude
$n_{\rm g}^{\rm rf}=0$--3 electrons, as well as the qubit DC-bias: $n_{{\rm g}0}$ and
$\phi$. In Fig. \ref{fig:PiFlux}  we present the reflection phase $\arg
(\Gamma)$ and magnitude $|\Gamma|$ measured at $f_{\rm rf}=4$ GHz on the
plane spanned by $n_{\rm g}^{\rm rf}$ and $n_{{\rm g}0}$. The Josephson capacitance
$C_{\mathrm{eff}} \sim - \arg (\Gamma) C_S^{3/2} Z_0/(2 \sqrt{L})$
deduced from the data of Fig. \ref{fig:PiFlux} has been given in
Ref. \cite{LZprl}\footnote{Full circuit analysis was employed in the
evaluation of the reported effective capacitance.}. We observe in
Fig. \ref{fig:PiFlux} a clear interference pattern whose main features
confirm the coherent LZ tunneling picture: 1) There is an onset of
the interference speckles, where the rf drive just reaches the
avoided crossing (charge degeneracy), $n_{{\rm g}0}\pm n_{\rm g}^{\rm rf} = n_{\rm g,deg}=$ odd integer, with a linear dependence between $n_{{\rm g}0}$  and the
AC drive amplitude. Additionally we have shown in  Ref. \cite{LZprl}
that 2) the density of the dots is proportional to $1/f_{\rm rf}$ in the
direction of $n_{{\rm g}0} $ as well as $n_{\rm g}^{\rm rf}$, and 3) the pattern
loses its contrast below a certain drive frequency, in this circuit
around $f_{\rm rf} \sim 2$~GHz.

 In Fig.~\ref{fig:PiFlux}a at charge degeneracy, we observe a clear sequence
of blue dots that signifies a similar response as in the ground
state. These dots correspond to the destruction of tunneling due to
periodic rf-drive of our two level system. Note also that these
locations correspond to a large reflection magnitude in Fig.~\ref{fig:PiFlux}b. In fact, the reflection is enhanced compared with
the undriven level, which indicates transfer of energy from the
microwave drive, via the qubit, to the resonator. Apparently, the
destruction of tunneling prevents the microwave energy from being
deposited to the two-level system, and the `extra' energy is dumped
out at both drive frequencies.

\begin{figure}
\begin{center}
\includegraphics[width=8.0cm]{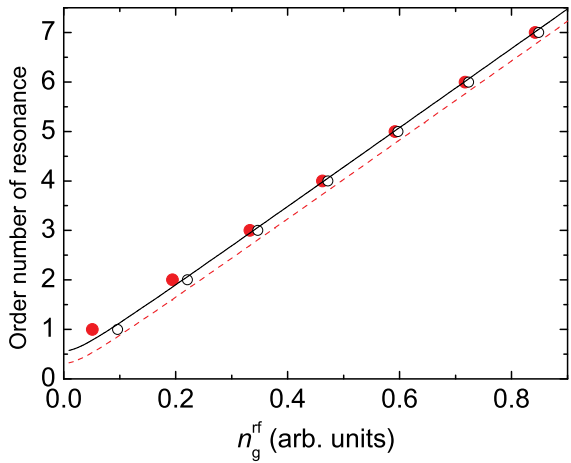} \caption{Order number of
destructive interference fringes at charge degeneracy ($n_{{\rm g}0}=1$)
as a function of the microwave drive amplitude $n_{\rm g}^{\rm rf}$ for phase
bias $\phi=0$. For comparison, the  dynamical phase calculated from
Eq.~(\ref{dynamic}) is given by the solid and dashed curves with and
without Stokes phase, respectively. The zeroes of Bessel function
$J_0$ are given by the open circles. For details, see
text.\label{fig:Stokes} }
\end{center}
\end{figure}

The order number of the destructive interference dots in Fig.
\ref{fig:PiFlux} is plotted in Fig. \ref{fig:Stokes}. At large
rf-drives, the interference dots display a linear dependence on the
drive amplitude. The linear dependence is a sign of asymptotic
behavior that can be obtained either from Eq. (\ref{phasecondition})
or from Eq. (\ref{bessel}).  Starting from Eq.
(\ref{phasecondition}), we may evaluate $\varphi_{\rm L,R}$ under a
drive of $\Delta E =4E_{\rm c} n_{\rm g}^{\rm rf}\sin(\omega_{\rm rf} t)$. This yields
$$\varphi_{\rm L,R}= -\frac{8 E_{\rm c} n_{\rm g}^{\rm rf}}{\hbar \omega_{\rm rf}} \mp m \pi
+ O\left(\frac{\hbar \omega_{\rm rf}}{4E_{\rm c} n_{\rm g}^{\rm rf}}\right),$$ where $m$
indexes the fringe number, and $\ell_{\rm R} = \ell_{\rm L} - m$, where $\ell_{\rm L,R}$ are the integers from Eq.~(\ref{phasecondition}).
Thus, we find
\begin{equation}\label{fit}
n_{\rm g}^{\rm rf}= \frac{\hbar\omega_{\rm rf}}{8E_{\rm c}}
\left[2\pi(\ell_{\rm L}-\frac{m+1}{2})-2\tilde{\varphi}_S\right]
\end{equation}
(note that the order of resonance grows with $\ell_{\rm L}$, and Eq.~(\ref{fit}) holds, of course, only for sufficiently large $\ell_{\rm L}$, when it gives $n_{\rm g}^{\rm rf}\leq0$).
For the destructive interference at charge degeneracy in the sudden
limit this yields $n_{\rm g}^{\rm rf}= ({2\pi\hbar
\omega_{\rm rf}}/{8E_{\rm c}})\left(\ell_{\rm L}-\frac{1}{4}\right)$. By
fitting Eq.~(\ref{fit}) to three/four highest drive points of the data in Fig.
\ref{fig:Stokes} we get $\varphi_S/\pi =0.30 \pm 0.03$, which is close to
the theoretical expectation of 0.25. However,
there is always an additional contribution from the dynamical phase
picked up near the degeneracy point where the bands are curving from linear.
This contribution was estimated to be about 1/3 from the expected Stokes phase.
This correction will drop the vertical offset, and reduce the Stokes phase down
to $0.20 \pi$. Hence, taking all the uncertainties into account, our final estimate for the Stokes phase
becomes $(0.2  \pm 0.05) \pi$ in the sudden limit.

A fit using the plain dynamic phase of Eq.~(\ref{dynamic}), and the
two level approximation for the adiabatic energy levels, is given by
the dashed curve in Fig. \ref{fig:Stokes}. The $x$-scale of the
calculated curve has been adjusted to match the experimental results
at large drive amplitudes. The solid curve, on the other hand, is
the dynamical phase corrected using the calculated Stokes phase in the sudden
limit. The curve agrees with data except at the lowest points where a
possible inaccuracy in the asymmetry parameter has its strongest
influence. The open circles display the result of Eq.~(\ref{bessel}),
\textit{i.e.} the zeroes of $J_0$, which are also seen to coincide
well with the data.

The Stokes phase is related to the non-adiabatic geometric phase,
the Aharonov-Anandan phase \cite{AAphase,Kayanuma97}. It is rather straightforward to show
that
\begin{equation}\label{AA-phase}
    \Phi_{\rm geom}=2(1-P_{LZ})\tilde{\varphi}_S+\pi(1-m)+\pi  P_{LZ}(2m+2\ell_{\rm L}-1).
\end{equation}
From this relation one gets at the charge degeneracy ($m=0$):
$\Phi_{\rm geom}=(1-P_{LZ})(2\tilde{\varphi}_S-\pi(2\ell_{\rm L}-1))$, which is
nearly the same relation as what we derived for the asymptotic fit
in Fig.~\ref{fig:Stokes}, cf.~Eq.~(\ref{fit}). Consequently, the geometric Aharonov-Anandan
phase could be approximately determined by taking the measured drive amplitude at the point of destructive interference and multiplying the
result by $-(8E_{\rm c}/\hbar\omega_{\rm rf})(1-P_{LZ})$. For the data in Fig. \ref{fig:Stokes} at 4 GHz,
we get a nearly constant value (0.20--0.21)~$\times 2\pi$
at drive amplitudes around $n_{\rm g}^{\rm rf}=0.5$--1.

\section{Artificial molecule}

In the inductive SET, LSET, of Fig. \ref{fig:Setup}b,
a stronger coupling between the qubit and the resonator can be
achieved than in the Cooper-pair-box configuration. This leads to more involved multiphonon phenomena which
are reminiscent of the transitions in diatomic
molecules. These transitions can also be viewed as interference
effects, and it is convenient and instructive to describe them as interference in the phase space \cite{Wheeler,Schleich}.

In order to understand the molecular analogy, we consider the single-Cooper-pair transistor
as a two-state system. The two states are the analog of two electronic states in a diatomic molecule.
The transistor is coupled in parallel with an $LC$
oscillator. The $LC$ oscillator is the analog of nuclear vibrations in the molecule.
Because of the coupling between the two-state system and the oscillator, a transition between the electronic levels is often accompanied by a change of the vibrational state. Such transitions, where both vibrational and electronic quantum numbers change simultaneously, are known as {\em vibronic}.

For quantitative analysis, we write the Hamiltonian in the basis of two relevant charge states of the qubit:
\begin{eqnarray}
H(\Phi,q)= \frac{1}{2}\left(\sigma_zE_{\rm
el}-\sigma_xE_{\rm J}\cos\frac{\pi\Phi}{\Phi_0}
+\sigma_yE_{\rm J}d\sin\frac{\pi\Phi}{\Phi_0}\right) +\frac{q^2}{2C}
+\frac{(\Phi-\Phi_{\rm b})^2}{2L}. \label{e.Hscowei}\end{eqnarray} Here
the flux $\Phi$ and the charge $q$ in the $LC$ oscillator are
conjugate variables, and $(\sigma_x,\sigma_y ,\sigma_z)$ are the Pauli matrices. The
capacitive energy $E_{\rm el} =2e^2(n_{{\rm g}0}-1)/C_\Sigma$ can be
controlled by the gate voltage. Another control parameter is the
flux bias $\Phi_{\rm b}$ through the loop containing SCPT and the
inductor.
%$E_{{\rm J}0}=E_{{\rm J}1}+E_{{\rm J}1}$ is the total Josephson energy of
%the SCPT and $d=(E_{{\rm J}1}-E_{{\rm J}1})/E_{{\rm J}0}$ is the asymmetry, and
%$\Phi_0=h/2e$ is the flux quantum.

In the following we concentrate on the limit, where the oscillator
frequency $1/\sqrt{LC}$ is much lower than the qubit level
difference. This means that all changes in the  qubit system are
much faster than in the oscillator system. Therefore, one can
diagonalize the qubit part separately from the oscillator. The two
energies of the qubit are obtained by diagonalizing the $2\times2$ matrix part of Eq.\ (\ref{e.Hscowei}):
\begin{eqnarray}
E=\pm \frac{1}{2}\sqrt{E_{\rm
el}^2+E_{\rm J}^2\cos^2\frac{\pi\Phi}{\Phi_0}+E_{\rm J}^2d^2\sin^2\frac{\pi\Phi}{\Phi_0}} \,.
\label{e.equbit}\end{eqnarray}
This form is valid when the Josephson coupling is a small perturbation compared to the capacitive energy $E_{\rm el}$.
It is also possible to consider the general case of an arbitrary ratio of the Josephson coupling to $E_{\rm el}$, in which case the energies of the two lowest states are given by Mathieu characteristics. This means that the dependence of the eigenenergies on $\Phi$ would be more complicated than in Eq.~(\ref{e.equbit}) but otherwise the following analysis remains intact.

\begin{figure}[tbp]
   \centering
   \includegraphics[width=0.8\linewidth]{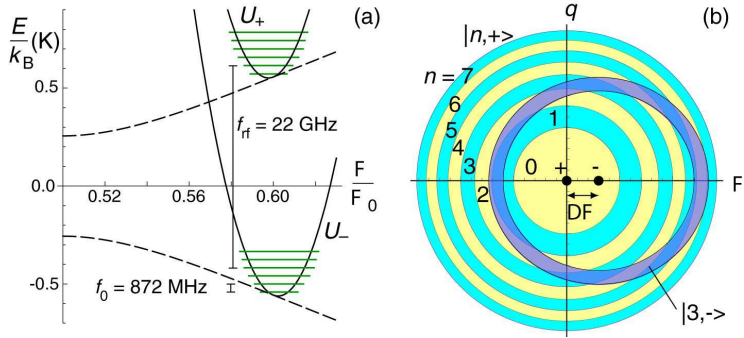} % requires the graphicx package
   \caption{The potentials and phase-space picture of vibronic
   transitions. (a) The potentials $U_\pm$ (solid curves)
   for vibrations correspond to two qubit states $\pm$. The qubit
   energy (\ref{e.equbit}) is given by dashed lines. The energies of six lowest vibrational states corresponding to both qubit states are drawn. The bars represent energy shifts
   induced by high and low frequency drives.  (b) The overlap of
   $|+,n\rangle$ states with $n=0,1, \ldots, 7$ vibrational quanta
   with  state $|-,3\rangle$ in the phase space.  The dots show the
   centers of the ellipses corresponding to qubit states $\pm$, which
   are displaced from each other by $\Delta\Phi$.}
   \label{f.oscpot}
\end{figure}

Considering now the oscillator, we can think about the capacitive and inductive terms in the Hamiltonian as the kinetic and potential energies. In addition to the inductive potential, there is a
potential arising from the qubit, since the qubit  energy
(\ref{e.equbit}) depends on the flux $\Phi$. Thus there are two
potential curves for the oscillations
\begin{eqnarray}
U_\pm(\Phi)=\pm \frac{1}{2}\sqrt{E_{\rm
el}^2+E_{\rm J}^2\cos^2\frac{\pi\Phi}{\Phi_0}+E_{\rm J}^2d^2\sin^2\frac{\pi\Phi}{\Phi_0}}
+\frac{(\Phi-\Phi_{\rm b})^2}{2L}.
\label{e.Hscoweiow}\end{eqnarray} Such
a potential is illustrated in Fig.\ \ref{f.oscpot}a. The different slopes of the qubit
   energies give rise to a relative shift of the minima of the oscillator
   potentials. The difference in the curvatures of the qubit energies
   shifts the vibrational frequencies. Figure \ref{f.oscpot}a, which is drawn to scale with realistic parameters of our circuit, shows that vibronic transitions can be induced by microwave radiation.
%%%%%%%%%%%%%%%%%%%%%%%%%%%%%%%%%%%%%%%%%
%%%%%%%%%%%%%%%%%%%%%%%%%%%%%%%%%%%%%%%%%

The intensities of vibronic  transitions are traditionally analyzed
in terms of the Franck-Condon principle, and this analysis turns out
to be useful for understanding of our data. In its classical form, the Franck-Condon principle says
that transitions are possible between vibrational states, the trajectories of which intersect in phase space,
and they are most intense between states with
coincident turning points. This is based on the idea, already
stated, that the transitions between electronic states are fast and
the vibrations are slow, so that the vibrational coordinates, here
the flux $\Phi$ and the charge $q$, have no time to change during the transition. In
addition, $\Phi(t)$ is slowest near the turning points, and therefore
transitions there are most likely.  In the following we discuss how
this picture can be extended by  semiclassical analysis in phase
space, as shown in Refs.~\cite{Wheeler,Schleich}, and how this interpretation is related to the data obtained.

The phase space formed by the coordinate $\Phi$ and the canonical
momentum $q$ is illustrated in Fig.\ \ref{f.oscpot}b. The
vibrational states correspond to elliptic rings. By scaling the
coordinates properly, a state with $n$ vibrational quanta in a
harmonic oscillator can be associated with a circular ring between
radii $\sqrt{n}$ and $\sqrt{n+1}$ (Planck-Bohr-Sommerfeld bands~\cite{Schleich}). This is the case for the upper
qubit state in Fig.\ \ref{f.oscpot}b. Due to a different resonance
frequency, the vibrational states corresponding to the lower qubit
state then appear as slightly squeezed in the $q$ direction.  The shift of
the minima of the potentials $U_\pm$ (Fig.~\ref{f.oscpot}a)
appears in the phase space so that the ellipses corresponding to the
states $|+,n\rangle$, where $+$ means the upper qubit state,  have
different centers than the ellipses corresponding to the states
$|-,n\rangle$. Vibronic transitions can occur only between states
whose ellipses overlap in the phase space. In the example of Fig.\
\ref{f.oscpot}b, the state $|-,3\rangle$ has overlap only with
states $|+,n\rangle$ with $1\leq n\leq 6$. In general, the transition rates are determined by the area of the intersection of the respective bands. This area is largest when the ellipses
touch each other tangentially, i.e., when the turning points coincide, and this
leads to the classical version of the Franck-Condon principle. In the present case though, the
overlap consists of two crossings of the ellipses, and one has to add the transition amplitudes taking into account the relative phase accumulated between the crossings. The matrix element of the transitions,  $|\langle+,n|-,m\rangle|^2$, is given by the formula~\cite{Wheeler,Schleich}
\begin{eqnarray}
P_{nm}=\frac{4 S_{nm}}{h}\cos^2\frac{A_{nm}}\hbar
\label{e.wheelerf}\end{eqnarray}
Here $S_{nm}$ is the area of one
of the crossing of the elliptic rings, which sets the maximum probability
of the transitions. The phase in the phase factor is determined by the area $A_{nm}$ between the two
alternative paths between the crossing points. Eq.~(\ref{e.wheelerf}) reproduces well the results of full quantum
calculation except near the classical turning points (where the two crossing areas merge). In the case of Fig.\ \ref{f.oscpot} the oscillations are nearly harmonic and the relative difference of frequencies is small.
 When, in addition, the
difference in the vibrational quanta is small,  $|n-m|\ll n$, the
probability $P_{nm}$ scales as $J_{n-m}^2(2b\sqrt{n})$, where
$b=\sqrt{\omega C/2\hbar}\Delta \Phi$ is the dimensionless separation of
the potential minima. Owing to the parabolic bands in the region
near the minimum gap, $b \propto (\Phi_{\rm b}/\Phi_0-1/2)$, when
$|\Phi_{\rm b}/\Phi_0-1/2| \ll 1$.

\begin{figure}
\begin{center}
\begin{minipage}{12pc}
\includegraphics[width=12pc]{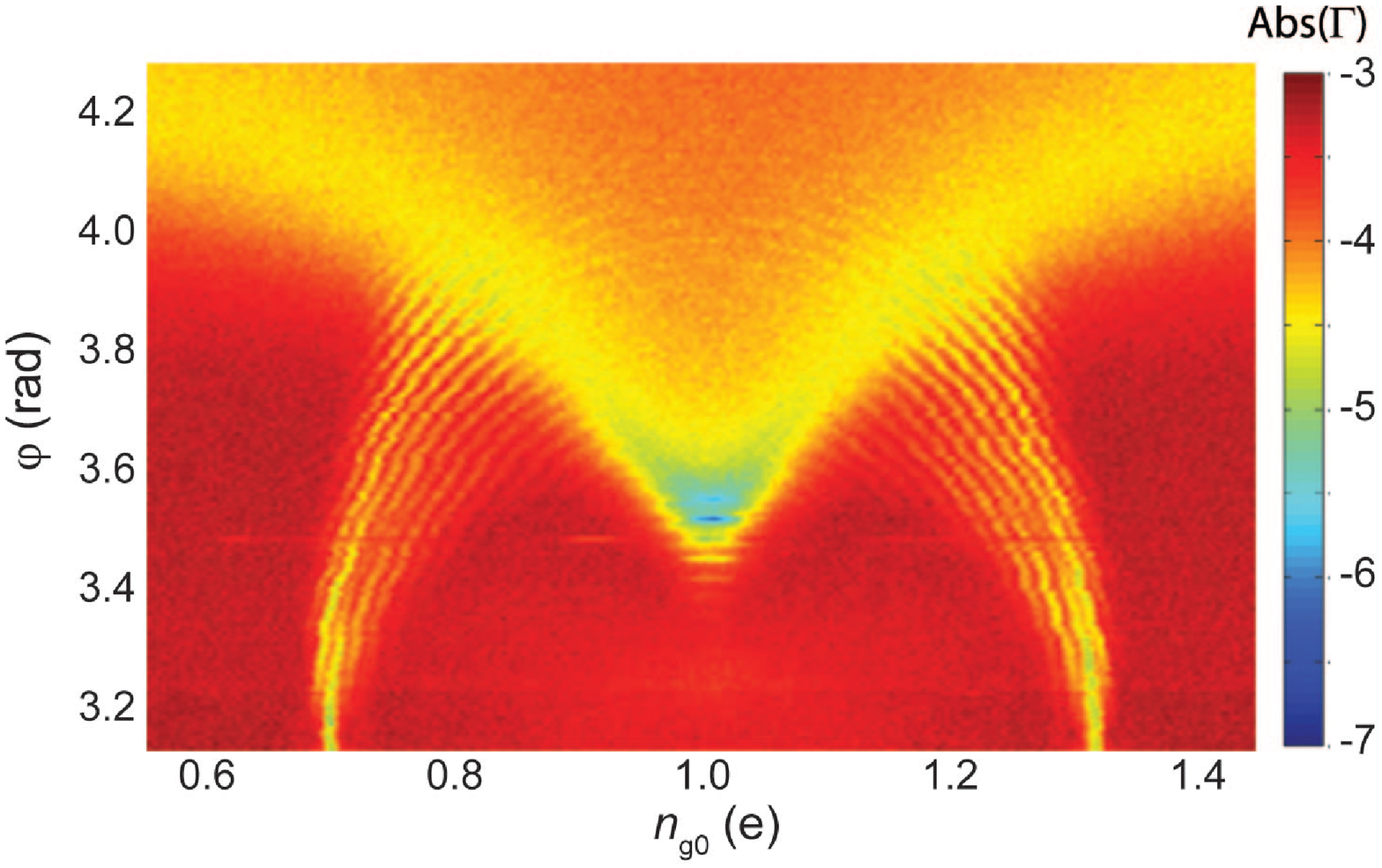} \caption{Measured amplitude $|\Gamma|$ of the reflection coefficient in the bias plane of gate charge $n_{{\rm g}0}$ and
phase $\phi=2\pi \Phi_{\rm b}/\Phi_0$. Several concentric circular fringes are visible below
the V-shaped light region. Out of them only the pure electronic transition ($n=m$) remains visible at $\phi=\pi$, and corresponds qubit level separation at $f_{\rm rf} =\Delta E/h=22$ GHz. The color bar gives the scale for the
magnitude of the reflection coefficient in dB.
%: $P_{\rm rf} = +15$ dBm
%(at the top of the cryostat) and
The measurement power was $P_{\rm 0} = -129$ dBm referred to the
coupling capacitor.
\label{fig:fringes}}
\end{minipage}\hspace{2pc}%
%%%%%%%%%%%%%%%%%%%%%%%%%%%%%%%%%%%%%%%%%
%%%%%%%%%%%%%%%%%%%%%%%%%%%%%%%%%%%%%%%%%
\begin{minipage}{12pc}
\includegraphics[width=12pc]{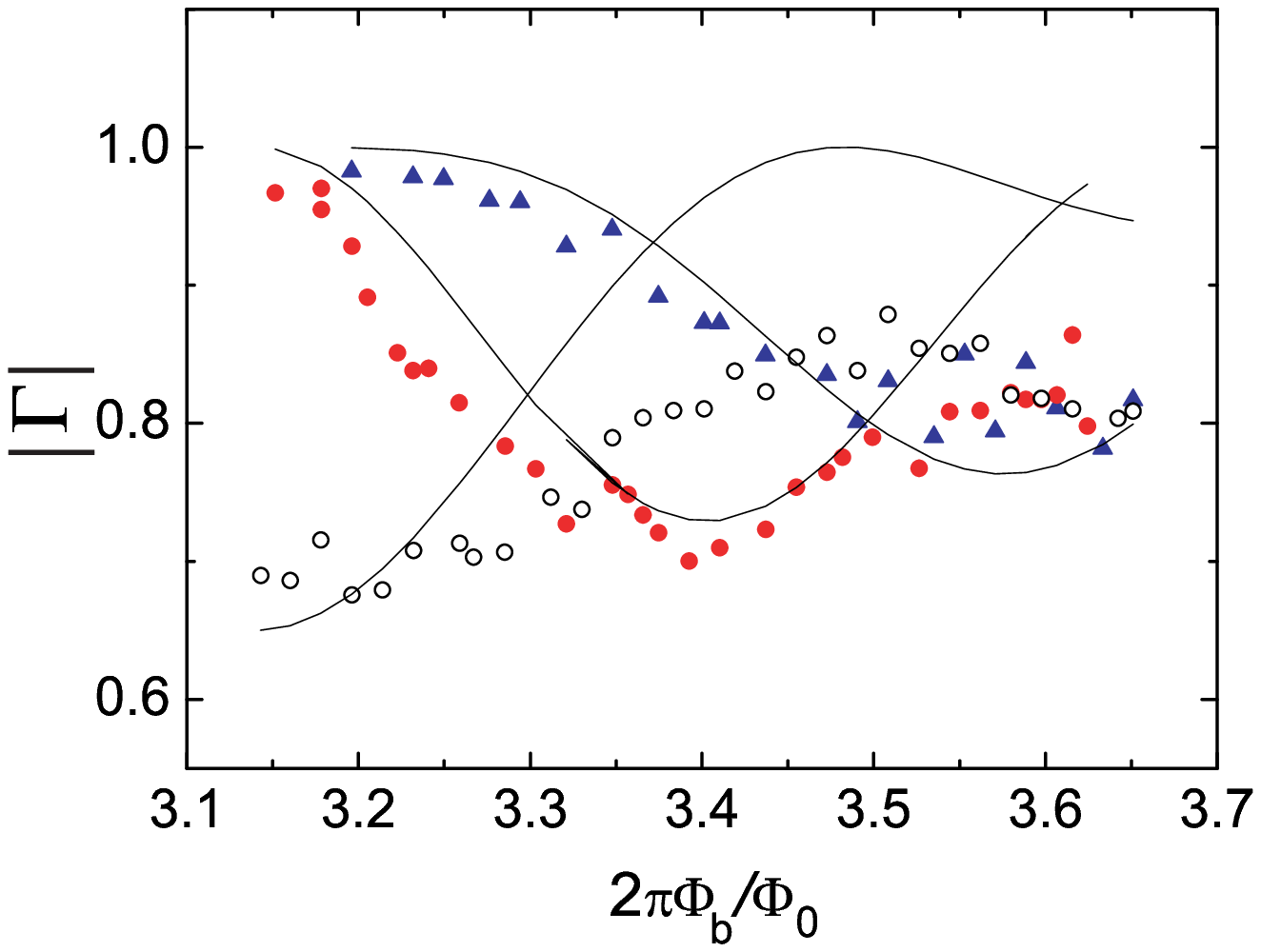} \caption{ Variation
of reflection coefficient along three fringes in
Fig. \ref{fig:fringes}. They correspond to $n-m=0$, -1, and -2 and are denoted
by $\circ$, $\bullet$, and $\blacktriangle$, respectively. The
fringe behavior is fitted to Bessel functions $J_0^2$, $J_1^2$, and
$J_2^2$. For details, see text.  \label{fig:bessel}}
\end{minipage}\hspace{2pc}%
\end{center}
\end{figure}

The measurements on the artificial molecule are made by analyzing
the reflection of microwaves at a frequency $f_0$ close to the resonance
frequency $f_{LC}$ of the circuit. In order to induce vibronic
transitions, a microwave excitation at frequency $f _{\rm rf}$
was used. The resonance condition for vibronic transition between states $\ket{-,m}$ and $\ket{+,n}$ is
\begin{eqnarray}
f_{\rm rf} \approx  \frac{\Delta E}{h} + (n-m) f_{LC},
\end{eqnarray}
where $\Delta E$ is the qubit energy splitting. A measurement of reflection coefficient $\Gamma$ in the
bias plane $(n_{\rm g},\Phi_{\rm b})$ is shown in Fig.~\ref{fig:fringes}.
%%%%%%%%%%%%%%%%%%%%%%%%%%%%%%%%%%%%%%%%%
%%%%%%%%%%%%%%%%%%%%%%%%%%%%%%%%%%%%%%%%%
The vibronic transitions are seen as fringes, making half-circles in
the bias plane. The fringes are  located within a crescent-shaped area. The language of interference in phase space allows one to easily understand, for instance, the shape of this area, the positions of the fringes and the modulation along the fringes.
At $\Phi_{\rm b}=\Phi_0/2$ only the pure electronic
transition is seen.  This agrees with the analysis above since there
the displacement of the oscillation minima vanishes. With increasing
$\Phi_{\rm b}$ more fringes appear, which is in agreement with the growing
separation of the minima, and the range of the fringes at a given $\Phi_{\rm b}$ can be determined as in Fig.~\ref{f.oscpot}b.
In addition, the above picture of enhanced
probability of the extremal trajectories is reflected in the shape
of the absorption pattern in Fig. \ref{fig:fringes}, where the edges
of the fan-like structure are rather clearly expressed.

The strength of the fringes in the reflection measurement depends on two separate factors.
One is the transition probability to the upper qubit state. In large part of the bias plane (below the the
V-shaped light region in Fig.~\ref{fig:fringes}) the averaged frequency of being part time in the upper and part time in the lower qubit state matches better to the measuring frequency than being in the lower state only. This leads to absorption which  is approximately proportional to the transition rate to the upper qubit state. The other factor is the heating or cooling effect caused by the high-frequency $f_{\rm rf}$ radiation on the vibrations. For a vibronic transition with $n >m$  (i.e. $f_{\rm rf} > \Delta E/h$), the resonance
deposits energy into the resonator causing heating of the vibrations. This leads to decreased absorption of the measurement wave. In the opposite case
$n <m$, the absorption of a high frequency photon cools the
oscillator by reducing its quanta which leads to  increased absorption of the measurement wave.
This asymmetry of the fringes is clearly visible in Fig.~\ref{fig:fringes}, where the cooling fringes at larger radii are stronger than the heating fringes at smaller radii.

Numerical simulations of the artificial molecule has been done in Ref. \cite{Gunnarsson}.
These are based  on Bloch equations describing the qubit and classical equations describing the circuit. They reproduce well the observed vibronic spectrum, both the matrix elements and the heating/cooling asymmetry. Here we present a simplified analysis by  comparing three of the fringes with the transition probability $P_{nm}=J_{n-m}^2(2b\sqrt{n})$, neglecting the heating/cooling effect.
Fig.~\ref{fig:bessel} displays the measured reflection magnitude
along fringes with $n-m=0$, -1, and -2 as a function of the flux bias of
the qubit. Squared Bessel functions $J_0^2$, $J_1^2$, and $J_2^2$
have been fitted to the data. For the $x$-scale argument we have taken
${\rm const} \times (\Phi_{\rm b}/\Phi_0-1/2)$ although this approximation will
be rather crude when $\Phi_{\rm b}/\Phi_0 > 0.54 $. Nevertheless, the comparison
can be employed to look for the presence of basic interference
phenomena. The agreement between the data and the fits is quite good, though the modulation of
the pure electronic fringe is a bit weaker in the measurement than
given by $J_0^2$.

The reason for smearing of the Bessel modulation is the variation of
the number of quanta in the resonator. Assuming the resonator  is in a coherent state, it is a superposition of the number
states and this should be taken into account in the analysis of the
interference fringes. This diminishes the sharpest features of the interference, but does not remove it.  Consequently, we may conclude that the dip at
$\Phi_{\rm b}/\Phi_0 \approx 0.54$  in the fringe magnitude corresponding
to the pure electronic transition, is due to destructive
interference in phase space in formula Eq.~(\ref{e.wheelerf}).

In summary, we have discussed interband transitions in
superconducting Cooper-pair boxes and charge-phase qubits and argued,
how interference effects in various forms can be found to underlie
the observed phenomena.

\section*{Acknowledgments}
Fruitful discussions with M.~Feigelman, T.~Heikkil\"a, F.~Hekking,
and M.~Paalanen are gratefully acknowledged.

%\bibliography{korotkov}

\end{document}